\documentclass[12pt]{iopart}

\def \->{\Rightarrow}
\newcommand{\del}[2]{\frac{\partial{#1}}{\partial{#2}}}
\newcommand{\mixdel}[3]{\frac{\partial^2{#1}}{\partial{#2}\partial{#3}}}
\newcommand{\ddel}[2]{\frac{\partial^2{#1}}{\partial{#2}^2}}
\newcommand{\be}{\begin{equation}}
\newcommand{\ee}{\end{equation}}
\newcommand{\seq}{\stackrel{\Sigma}{=}}
\newtheorem{prop}{Proposition}

\newtheorem{lemma}{Lemma}

\begin{document}

\title[On isotropic cylindrically symmetric stellar models]
{On isotropic cylindrically symmetric stellar models}
\author{Brien C Nolan and Louise V Nolan}
\address{School of Mathematical Sciences, Dublin City University, Glasnevin, Dublin 9,
Ireland} \eads{brien.nolan@dcu.ie, louise.nolan3@mail.dcu.ie}
\begin{abstract}
We attempt to match the most general cylindrically symmetric vacuum space-time with a
Robertson-Walker interior. The matching conditions show that the interior must be dust filled and
that the boundary must be comoving. Further, we show that the vacuum region must be polarized.
Imposing the condition that there are no trapped cylinders on an initial time slice, we can apply a
result of Thorne's and show that trapped cylinders never evolve. This results in a simplified line
element which we prove to be incompatible with the dust interior. This result demonstrates the
impossibility of the existence of an isotropic cylindrically symmetric star (or even a star which
has a cylindrically symmetric portion). We investigate the problem from a different perspective by
looking at the expansion scalars of invariant null geodesic congruences and, applying to the
cylindrical case, the result that the product of the signs of the expansion scalars must be
continuous across the boundary. The result may also be understood in relation to recent results
about the impossibility of the static axially symmetric analogue of the Einstein-Straus model.
\end{abstract}
 \maketitle

\section{Introduction}
In this article, we study non-spherical gravitational collapse. This is an area where not
much is currently known in the highly non-spherical regime, where perturbative studies
are not appropriate: this is in contrast to the near-spherical case. To study this
physical situation, mathematical assumptions must be made. Ideally, one would like to
study dynamical axially symmetric situations, leading to PDE's in 2+1 dimensions. To
attempt to capture the flavour of possible outcomes and to pose tractable problems, a
further symmetry is added (translational symmetry along an axis), leading to cylindrical
symmetry. This models highly prolate collapse and as mentioned below, can be used to
construct axially symmetric models by the addition of hemispherical caps to a finite
cylindrical object.\\Cylindrically symmetric space-times in General Relativity are
idealized models, but they provide insight into non-spherical gravitational collapse and
the non-linearity of the field equations. Key results in this field include the
following: Thorne showed that horizons cannot evolve in the vacuum region surrounding a
collapsing infinite cylinder \cite{thorne4}. Berger, Chrusciel and Moncrief proved that
asymptotic flatness, energy conditions and cylindrical symmetry exclude the existence of
compact trapped surfaces \cite{berg1}. Apostolatos and Thorne proved that even an
infinitesimal amount of rotation can halt the collapse of an infinite cylindrical dust
shell \cite{thorne5}. More recent work describes the cylindrically symmetric collapse of
an infinite null dust shell \cite{echev1}, non-rotating, infinite dust cylinders
\cite{singh1},\cite{nakao1}, counter-rotating dust shells
\cite{gon1},\cite{nolan1},\cite{per1} and self-similar scalar field \cite{wang1}. These,
and other results furnish us with a clearer picture of non-spherical collapse in which
gravitational
radiation, angular momentum and critical phenomena play an important role.\\
In many instances these results are in sharp contrast to results for the corresponding
spherically symmetric model, for example Berger et al's strong cosmic censorship result
\cite{berg1} and Thorne's \cite{thorne4} and Golcalves' \cite{gon2} results ruling out
certain types of horizon. So with a view to comparing and contrasting cylindrical and
spherical collapse, we investigate a cylindrical version of the ``standard'' model of
spherical collapse, the Oppenheimer-Snyder model \cite{opp}. However unlike that model we
do not use the \textit{a priori} assumption that the interior comprises pressureless
dust. In this paper we apply the standard matching techniques, without any conditions of
staticity, to the cylindrically symmetric case to obtain a general result about the
evolution of cylindrically symmetric
objects in a vacuum space-time.\\
As a preliminary we examine a general matching of a space-time $(\mathcal{V}^-,g^-)$ to a
vacuum space-time $(\mathcal{V}^+,g^+)$, with the matching condition \cite{mars0},
\begin{eqnarray}\label{darmois}
[T_{\mu\nu}]n^{\nu}=\left(T^+_{\mu\nu}-T^-_{\mu\nu}\right)n^{\nu}=0,
\end{eqnarray}
where $T^\pm_{\mu\nu}$ is the energy momentum tensor in $\mathcal{V}^\pm$ respectively.
The conditions (\ref{darmois}) are sometimes known as the Israel junction conditions and
follow from the standard (Darmois) matching conditions of General Relativity - continuity
of the first and second fundamental forms - which are used throughout. For vacuum
$\mathcal{V}^+$, (\ref{darmois}) implies that
\begin{eqnarray}\label{m1-condition}
T^-_{\mu\nu}n^{\nu}=0,
\end{eqnarray}
on the matching hypersurface $\Sigma$. Assuming that the energy momentum tensor of
$(\mathcal{V}^-,g^-)$ is that of a perfect fluid, we have
\begin{eqnarray*}\label{emtensor}
T^-_{\mu\nu}=(p+\rho)u_{\mu}u_{\nu}+pg_{\mu\nu},
\end{eqnarray*}
where $u_{\mu}$ is a unit future pointing time-like vector. Then (\ref{m1-condition})
becomes
\begin{eqnarray}\label{m2-condition}
pn_{\mu}+(p+\rho)n_{\nu}u^{\nu}u_{\mu}=0.
\end{eqnarray}
If we invoke the weak energy condition, $\rho\geq{0}$ and $\rho+p\geq{0}$, and require
$\rho\neq{0}$ to avoid a trivial case, then (\ref{m2-condition}) implies that $p=0$ on
$\Sigma$ and $u_{\nu}n^{\nu}=0$. In other words, matching with vacuum can only be done
across a time-like hypersurface on which the pressure vanishes. We are considering the
case where $(\mathcal{V}^-,g^-)$ is a Robertson-Walker (RW) space-time. Since the
pressure of such a space-time is homogeneous, this implies that the pressure must vanish
everywhere. So we will consider a RW interior space-time $(\mathcal{V}^-,g^-)$ with
cylindrically symmetric line element in coordinates $\{t,\rho,x,\varphi\}$
\begin{eqnarray}\label{dustmetric}
ds^{2}_-&=&-dt^2+a^2(t)(d\rho^2+\Upsilon^2_{,\rho}(\rho,\epsilon)dx^2+\Upsilon^2(\rho,\epsilon)d\varphi^2),
\end{eqnarray}
where $a(t)$ is the scale factor and for collapsing dust
\[ a(t)=a_0|t|^{2/3},\quad t\in(-\infty,0],\]  and where $\Upsilon(\rho,\epsilon)$ satisfies
\begin{eqnarray}
\Upsilon(\rho,\epsilon)=\left\{ \begin{array}{ll}
             \sinh\rho,&\epsilon=-1,\\
             \rho,&\epsilon=0,\\
             \sin\rho,&\epsilon=+1,\end{array} \right.
\end{eqnarray}
and where $\epsilon$ is the curvature index so that $\epsilon=1,0,-1$ for closed, flat or
open RW models, respectively. We will match to a general cylindrically symmetric
unpolarized vacuum exterior space-time $(\mathcal{V}^+,g^+)$ given by the line element in
coordinates $\{T,R,Z,\Phi\}$ \cite{exactsol} (see also \cite{vera2},\cite{barnes1})
\begin{eqnarray}
\label{vacmetric}ds^{2}_+&=&{e}^{2(\gamma-\psi)}(-dT^2+dR^2)+
{e}^{2\psi}(dZ+\omega{d\Phi})^2+\alpha^2{e}^{-2\psi}d\Phi^2,
\end{eqnarray}and $\gamma$, $\psi$,
$\omega$ and $\alpha$ are functions of $T$ and $R$\footnote{The intention is that if the
matching were successful we would propose a global picture of the whole matched
space-time with the infinite RW cylinder truncated at two values of $z$ and hemispherical
caps inserted at these points $z=z_1$ and $z=z_2$ so that the RW portion corresponds to
$z_1<z<z_2$\cite{israel1}. Thus the vacuum region should include the axis beyond the
caps. This implies orthogonal transitivity of the isometry group and so (\ref{vacmetric})
applies.\cite{vera2}}. In \S\ref{sec:hypersurface} we describe the most general matching
hypersurface. In \S\ref{sec:induct} by assuming that $\omega(R,T)$ is analytic in $R$ and
$T$ on $\Sigma$ we show that $\omega=0$ everywhere, i.e.\ that an unpolarized vacuum
space-time cannot be matched to a dust interior. In \S\ref{sec:polargen} we show that, by
applying an argument due to Thorne and requiring that there are no trapped surfaces
initially, we can reduce to the case $\alpha(R,T)=R$. In \S\ref{sec:polar}, we show that
the matching of the (now simplified) exterior to the RW interior is impossible. We also
show that continuity of the metric alone rules out the existence of $\Sigma$. Thus one
cannot match the RW interior with the polarized vacuum exterior even if there is a
distributional shell of matter forming the boundary. This demonstrates the impossibility
of matching a RW interior to a cylindrically symmetric vacuum exterior.
\section{Matching hypersurface}\label{sec:hypersurface}
We begin with the following coordinate systems
\begin{eqnarray*}
\mbox{Interior}\ {\mathcal{V}^-}: x^{\mu}_-&=&\{t,\rho,x,\varphi\},\\
\mbox{Exterior}\ {\mathcal{V}^+}: x^{\mu}_+&=&\{T,R,Z,\Phi\},\\
\mbox{Matching hypersurface}\ {\Sigma}:{\xi}^{a}&=&\{\tau,z,\phi\}.
\end{eqnarray*}
We use the following conventions: $\mu,\nu,..=0,1,2,3$ and $a,b,..=1,2,3$, with prime and
overdot referring to differentiation with respect to $T$ and $t$ respectively. In order
to have a description of $\Sigma^{\pm}$, the hypersurface in $\mathcal{V}^\pm$
respectively ($\Sigma^{\pm}\subset\mathcal{V}^\pm$), in terms of the coordinates
$\{\tau,z,\phi\}$ we proceed as follows \cite{vera1}. A careful choice of coordinate
$\phi$, use of the commutativity of the Killing vectors in $\mathcal{V}^+$, observation
of the uniqueness of the generator of axial symmetry and a coordinate transformation
$(Z,\Phi)\rightarrow(\tilde{Z}(Z),\tilde{\Phi}(Z,\Phi))$ that does not alter the form of
the line element in $\mathcal{V^+}$ are all needed to arrive at the following description
of $\Sigma^{\pm}$
\begin{eqnarray}
\label{sig1}\Sigma^-&:&\{t=\tau,\ \rho=\rho_0,\ x=z,\ \varphi=\phi\},\\
\label{sig2}\Sigma^+&:&\{T=T_0(t),\ R=R_0(t),\ Z=z,\ \Phi=\phi\},
\end{eqnarray}
with the requirement $0<\rho_0$. In theory four different matchings of
$(\mathcal{V}^-,g^-)$ and $(\mathcal{V}^+,g^+)$ are possible, depending on the choice of
continuous normal $\vec{n}^{\pm}$ to $\Sigma^{\pm}$ in $\mathcal{V}^{\pm}$. However our
aim is to describe a space-time consisting of a RW interior and a vacuum exterior. For
$\vec{n}^{-}$, we choose the normal to point toward cylinders of increasing radius. For
$\vec{n}^{+}$, we want to do the same. This requires that $\vec{n}^{+}$ points towards
larger values of $\alpha$. The coordinate $R$ has yet to be specified and it may happen
that either $\alpha_{,R}>0$ or $\alpha_{,R}<0$. We assume further that the axis of the
vacuum space-time resides in the region removed to accommodate the RW portion. Thus
$\alpha_{,R}<0$ in $\mathcal{V}^{+}$ can only come about if $R$ decreases away from
$\Sigma^+$. Hence the $\alpha_{,R}<0$ case can be converted to the $\alpha_{,R}>0$ case
by a coordinate transformation of the form
\[R\rightarrow\hat{{R}}=R^*-R.\] Thus in the coordinates of (\ref{sig2}), we will assume that $\vec{n}^{+}$
points in the direction of increasing $R$. We will refer to this arrangement of
$\Sigma^{\pm}$ and $\vec{n}^{\pm}$ by saying that $\vec{n}$ points out of $\mathcal{V}^-$
and into $\mathcal{V}^+$.

\section{Reduction to the polarized case}\label{sec:induct}
We begin with the line element (\ref{vacmetric}). The tangent vectors,
$e^{\mu}_a=\frac{\partial{x^{\mu}}}{\partial{\xi^a}}$, to the hypersurface, $\Sigma$ are
\begin{eqnarray*}
e^{\mu}_1&=&\left(\frac{\partial{T}}{\partial{t}},\frac{\partial{R}}{\partial{t}},0,0\right),\\
e^{\mu}_2&=&(0,0,1,0),\\
e^{\mu}_3&=&(0,0,0,1),
\end{eqnarray*}
so that
\begin{eqnarray*}
\vec{{e}}_1&=&e^{\mu}_1\frac{\partial}{\partial{x^\mu}}=\frac{dT}{dt}\left(\frac{\partial}{\partial{T}}
+(R_0^\prime)\frac{\partial}{\partial{R}}\right).
\end{eqnarray*}
Thus there is a tangential derivative proportional to
\begin{equation*}
\frac{\partial}{\partial{T}}+(R_0^\prime)\frac{\partial}{\partial{R}},
\end{equation*}
which for convenience we will refer to as \textit{the} tangential derivative. (The other
tangential derivatives, $\partial_{\phi}$, $\partial_z$, are trivial in the sense that
they play no role in
the dynamics.)\\
We will require the following matching equations:
\begin{eqnarray}
\label{match11}h_{z\phi}^+\seq{h}_{z\phi}^-&\Leftrightarrow&\omega_{\Sigma}\seq{0},\\
\label{match12}h_{\tau\tau}^+\seq{h}_{\tau\tau}^-&\Rightarrow&\left(1-\left({R_0}^{\prime}\right)^2\right)>0,\\
\label{match13}K_{z\phi}^+\seq{K}_{z\phi}^-&\Leftrightarrow&\left((R_0^{\prime})\del{\omega}{T}+\del{\omega}{R}\right)\seq{0},
\end{eqnarray}
where $h_{ab}$ is the first fundamental form, $K_{ab}$ is the second fundamental form and
we use $\seq$ to indicate equality on $\Sigma$. We can take the tangential derivative of
$\omega$ and then evaluate it on $\Sigma$
\begin{equation*}
\left(\frac{\partial\omega}{\partial{T}}+
(R_0^\prime)\frac{\partial\omega}{\partial{R}}\right)\seq{0}.
\end{equation*}
By our matching condition (\ref{match13}) this implies
\begin{eqnarray*}
\frac{\partial\omega}{\partial{T}}(1-(R_0^\prime)^2)\seq{0}.
\end{eqnarray*}
Using the matching conditions (\ref{match12}) then gives the result
\begin{eqnarray}
\label{eq:ind1}\->\frac{\partial\omega}{\partial{T}}\seq{0},
\quad\frac{\partial\omega}{\partial{R}}\seq{0},
\end{eqnarray}
or equivalently
\begin{eqnarray}
\label{eq:ind2}\omega^{\left(1\right)}\seq{0},
\end{eqnarray}
where $\omega^{\left(k\right)}$ denotes all partial derivatives of $\omega$ of order $k$.\\
It is then straightforward to show that $\omega^{\left(2\right)}\seq{0}$ by considering
the field equation (\ref{eq:field}), which we can write in the form
\begin{eqnarray}
\label{eq:field0}\frac{\partial^2\omega}{\partial{T^2}}-\frac{\partial^2\omega}{\partial{R^2}}=f(\omega,\omega^{(1)}),\
\end{eqnarray}
where $f$ is some polynomial function satisfying $f(0,0)=0$, which by (\ref{match11}) and
(\ref{eq:ind2}) equals zero when evaluated at $\Sigma$. By taking tangential derivatives
of both of (\ref{eq:ind2}) and comparing with (\ref{eq:field0}), we obtain the result
$\omega^{\left(2\right)}=0$.

In like manner, we can show that if $\omega$ is $C^k,k\leq\infty$ on a neighbourhood of
$\Sigma$, then $\omega^{(k)}\seq0.$ We use an induction argument. If we assume that
$\omega^{\left(j\right)}\seq{0}$ for $0\leq{j}\leq{k}$ is true then by proving
$\omega^{\left(k+1\right)}\seq{0}$ is true and using (\ref{eq:ind2}) we have the desired
result.

To show that $\omega^{\left(k+1\right)}\seq{0}$ is true, we take the tangential
derivative of our assumed $\omega^{\left(k\right)}\seq{0}$. There are $(k+1)$ of these
tangential derivative equations, and they are of the form
\begin{eqnarray}
\label{eq:deriv1}\frac{\partial^{k+1}\omega}{\partial{T^{k+1}}}+{R}_0^{\prime}\frac{\partial^{k+1}\omega}{\partial{R}\partial{T^k}}&\seq&0,\\
\label{eq:deriv2}\frac{\partial^{k+1}\omega}{\partial{T^{k}}\partial{R}}+{R}_0^{\prime}\frac{\partial^{k+1}\omega}{\partial{R^2}\partial{T^{k-1}}}&\seq&0,\\
\label{eq:deriv3}\frac{\partial^{k+1}\omega}{\partial{T^{k-1}}\partial{R^2}}+{R}_0^{\prime}\frac{\partial^{k+1}\omega}{\partial{R^3}\partial{T^{k-2}}}&\seq&0\quad\mbox{etc.}\
\end{eqnarray}
Then we consider the field equation (\ref{eq:field0}). Taking successive partial
derivatives of (\ref{eq:field0}) we have
\[\frac{\partial^{k+1}\omega}{\partial{T^{k+1}}}-\frac{\partial^{k+1}\omega}{\partial{T^{k-1}}\partial{R^2}}=F(\omega,\cdots,\omega^{(k)}),\]
and the form of $f$ in (\ref{eq:field}) shows that $F(0,\cdots,0)=0$. But evaluated on
$\Sigma$ we know by assumption that
\[\omega^{\left(j\right)}\seq{0}\  \mbox{for}\ 0\leq{j}\leq{k},\]
and therefore
\[\frac{\partial^{k+1}\omega}{\partial{T^{k+1}}}-\frac{\partial^{k+1}\omega}{\partial{T^{k-1}}\partial{R^2}}\seq{0}.\]
This equation together with (\ref{eq:deriv1}) and (\ref{eq:deriv2}) gives the relation
\begin{eqnarray*}
\frac{\partial^{k+1}\omega}{\partial{T^{k}}\partial{R}}(1-({R}_0^{\prime})^2)\seq{0},
\end{eqnarray*}
and so by (\ref{match12})
\begin{eqnarray*}
\frac{\partial^{k+1}\omega}{\partial{T^{k}}\partial{R}}\seq{0}.
\end{eqnarray*}
Substituting this equation into the appropriate tangential equation shows, by a cascade
effect, each partial derivative of order $(k+1)$ to be zero when evaluated at $\Sigma$,
proving our assertion. We can then write down the following lemma.
\begin{lemma}
If $\omega$ is analytic on a neighbourhood $\Omega$ of $\Sigma$, then $\omega\equiv0$ on
$\Omega$.
\end{lemma}
\noindent{\bf Proof:} Let $(R_1,T_1)\in\Omega$. Then we can write
$R_1=R_0(T)+R_*,T_1=T+T_*$ for some numbers $R_*,T_*$ and where $(R_0(T),T)\in\Sigma$. By
analyticity, we can write
\begin{eqnarray}
\label{induct}\omega(R_0(T)+R_*,T+T_*)=\left[\sum_{n=1}^{\infty}\frac{1}{n!}\left(T_*\frac{\partial}{\partial{T}}+R_*\frac{\partial}{\partial{R}}\right)^n\omega\right]_{\Sigma}.
\end{eqnarray}
The result follows immediately.

Therefore, assuming that $\omega(R,T)$ is an analytic function we see that $\omega=0$ on
a neighbourhood of $\Sigma$ by using the matching conditions and the vacuum field
equations. So we have henceforth that $\omega=0$ and (\ref{vacmetric}) becomes
\begin{eqnarray}
\label{vacmetric1}ds^{2}_+&=&{e}^{2(\gamma-\psi)}(-dT^2+dR^2)+
{e}^{2\psi}dZ^2+\alpha^2{e}^{-2\psi}d\Phi^2.
\end{eqnarray}

\section{Further simplification of the vacuum line element}\label{sec:polargen}
We note that the general solution of the vacuum field equation
\begin{eqnarray}\label{eq:field1}
\ddel{\alpha}{R}-\ddel{\alpha}{T}&=&0,
\end{eqnarray}is
\begin{eqnarray}
\quad\label{eq:field2} \alpha(T,R)&=&F(U)+G(V),
\end{eqnarray}
where $U=T-R$ and $V=T+R$. Following Thorne \cite{thorne5} we define the character of
space-time with line element (\ref{vacmetric1}) at any event $p$ to be $D^{(+)}$ if
$\nabla\alpha$ is space-like and points away from the symmetry axis, $D^{(-)}$ if
$\nabla\alpha$ is space-like and points toward the symmetry axis, $D^{(0\uparrow)}$ if
$\nabla\alpha$ is time-like and points toward the future, and $D^{(0\downarrow)}$ if
$\nabla\alpha$ is time-like and points toward the past. So at any event $p$, the
character is
\begin{eqnarray*}
D^{(+)}&\Leftrightarrow&\del{F}{U}<0,\ \del{G}{V}>0,\\
D^{(-)}&\Leftrightarrow&\del{F}{U}>0,\ \del{G}{V}<0,\\
D^{(0\uparrow)}&\Leftrightarrow&\del{F}{U}>0,\ \del{G}{V}>0,\\
D^{(0\downarrow)}&\Leftrightarrow&\del{F}{U}<0,\ \del{G}{V}<0.
\end{eqnarray*}
It is straightforward to show that $D^{(+)}$ and $D^{(-)}$ at $p$ imply no trapped
cylinders at $p$, while $D^{(0\uparrow)}$ and $D^{(0\downarrow)}$ implies trapping. To
prove this we note that the standard line element (\ref{vacmetric1}) can be rewritten in
terms of null coordinates $U$ and $V$
\begin{eqnarray*}
ds^2_+&=&-{e}^{2(\gamma-\psi)}dUdV+ {e}^{2\psi}dZ^2+\alpha^2{e}^{-2\psi}d\Phi^2.
\end{eqnarray*}
The condition for a two-cylinder, $S$, of constant $T$ and $R$ to be untrapped is that
\begin{eqnarray*}
\theta_+\theta_-<0,
\end{eqnarray*}
where $\theta_{\pm}$ are respectively the expansions of the future-pointing outgoing and
ingoing null geodesics $l^{\mu}_{\pm}$ orthogonal to $S$, given by
\begin{eqnarray*}
l_+^{\mu}=a(V)e^{-2(\gamma-\psi)}\delta_U^{\mu},\\
l_-^{\mu}=b(U)e^{-2(\gamma-\psi)}\delta_V^{\mu},
\end{eqnarray*}
where $a(V)>0$ and $b(U)>0$. We find the expansions of these null geodesics to be
\begin{eqnarray}\label{app1}
\theta_+:=\nabla_{\mu}{l}^{\mu}_{+}=a(V)e^{-2(\gamma-\psi)}{\alpha_{,U}\over\alpha},\\
\label{app2}\theta_-:=\nabla_{\mu}{l}^{\mu}_{-}=b(U)e^{-2(\gamma-\psi)}{\alpha_{,V}\over\alpha},
\end{eqnarray} and we can write
\begin{eqnarray*}
\theta_+\theta_-=a(V)b(U)e^{-4(\gamma-\psi)}\frac{\alpha_{,V}\alpha_{,U}}{\alpha^2}=a(V)b(U)\frac{e^{-4(\gamma-\psi)}}{\alpha^2}\frac{dF}{dU}\frac{dG}{dV}.
\end{eqnarray*}
The result follows by inspection.\\
We require that initially there are no trapped surfaces
\[{D}^{(+)}\ \mbox{or}\ {D}^{(-)}\ \mbox{at}\ T=0.\]
However we note that in the absence of trapped cylinders
\begin{eqnarray*}
D^{(+)}&\Leftrightarrow&\del{\alpha}{R}>0,\\
D^{(-)}&\Leftrightarrow&\del{\alpha}{R}<0.
\end{eqnarray*}
These constraints, together with the assumptions of $\S2$ rule out ${D}^{(-)}$
\begin{eqnarray}\label{char1}\Rightarrow{D}^{(+)}\ \mbox{at} \
T=0.\end{eqnarray} Thorne \cite{thorne5} showed that in the vacuum region outside a
cylindrical shell of matter, with the constraint (\ref{char1}), the only possible
character change is
\[D^{(+)}\rightarrow{D}^{(0\downarrow)}.\]However this happens on an ingoing null
hypersurface which would intersect
$T=0$\begin{eqnarray}\label{char2}\Rightarrow{D}^{(0\downarrow)}\ \mbox{at}\ T=0\
\mbox{for}\ R>R_*,\end{eqnarray}where $R_*\geq{0}$. The contradiction between equations
(\ref{char1}) and (\ref{char2}) implies that
\[{D}^{(+)}\ \mbox{at} \
T=0\Rightarrow{D}^{(+)}\ \forall\ {T}\geq0.\] The argument also holds in the vacuum
region outside our cylindrical star.

Furthermore it has been shown \cite{thorne3} that in a space-time of character $D^{(+)}$
we can make a coordinate transformation
\begin{eqnarray}
\left(T,R\right)\rightarrow\left(\hat{T}(T,R),\hat{R}(T,R)\right),
\end{eqnarray} whereby $\alpha(R,T)$ becomes the new radial
variable $\hat{R}$. Therefore if our vacuum space-time does not contain trapped cylinders
initially and is not radially closed \cite{thorne5} ($D^{(+)}$ at $T=0$) we can use the
above results to describe the vacuum exterior space-time, $(\mathcal{V}^+,g^+)$, by
\begin{eqnarray}
ds^{2}_+&=&-{e}^{2(\gamma-\psi)}(dT^2-dR^2)+ {e}^{2\psi}dZ^2+R^2{e}^{-2\psi}d\Phi^2,
\end{eqnarray}
where we have rewritten $\hat{T}$ and $\hat{R}$ as $T$ and $R$ without confusion.

\section{Impossibility of the matching}\label{sec:polar}
Thus far, we have shown that the most general matching of a non-vacuum RW universe  with
a vacuum cylindrically symmetric space-time reduces to the  case where the RW universe is
dust-filled, the boundary is co-moving, the vacuum region is polarized and has character
$D^{(+)}$. In this section, we show that this matching configuration is impossible. More
generally, we show that metric matching alone rules out the matching of a collapsing RW
universe across a co-moving hypersurface with a polarized cylindrical vacuum space-time.
Thus even the insertion of a distributional shell of matter (which would arise in the
case where $K_{\mu\nu}$ is discontinuous) cannot yield a solution to the matching
problem.

The interior line element is \be
ds^{2}_-=-dt^2+a^2(t)(d\rho^2+\Upsilon^2_{,\rho}(\rho,\epsilon)dx^2+\Upsilon^2(\rho,\epsilon)d\varphi^2),\ee
and the exterior line element is \be
ds^2_+=-e^{2(\gamma-\psi)}(dT^2-dR^2)+e^{2\psi}dZ^2+R^2e^{-2\psi}d\Phi^2.
\label{vacmetric3}\ee By a collapsing RW universe, we mean one for which the scale factor
$a(t)$ decays to zero in finite time;
\[ \lim_{t\to 0^-}a(t)=0,\]
where by a time translation we have set the time of complete collapse to be at $t=0$. Of
course this includes the dust model considered above. Note that since we have dropped the
junction condition $[K_{\mu\nu}]=0$, the matching condition (\ref{darmois}) no longer
holds, and so we are not restricted to dust. Metric continuity across the comoving
hypersurface $\rho=\rho_0$ yields
\[ \Upsilon_{,\rho}(\rho,\epsilon)a(t) \seq \exp(\psi(R_0(T),T)),\]
where
\begin{eqnarray}
\left.\Upsilon_{,\rho}(\rho,\epsilon)\right|_{\Sigma}=\left\{ \begin{array}{ll}
             \cosh\rho_0,&\epsilon=-1,\\
             1,&\epsilon=0,\\
             \cos\rho_0,&\epsilon=+1.\end{array} \right.
\end{eqnarray}
We note that if $\rho_0=\frac{\pi}{2}$ in the case $\epsilon=+1$, then the matching
conditions are violated. So we rule out this case. Noting then that
$\left.\Upsilon_{,\rho}(\rho,\epsilon)\right|_{\Sigma}\neq{0}$, we immediately obtain \be
\lim_{T\to T_*} \psi(R_0(T),T)=-\infty,\label{mainlim}\ee where
\[ T_*:=\lim_{t\to 0^-}T_0(t),\]
where $T_0(t)$ is the solution of the metric matching condition
\[
\left(\frac{dT_0}{dt}\right)^2\seq{e}^{-2(\gamma-\psi)}(1-(R_0^\prime)^2)^{-1}.\] Now
$\psi$ satisfies the linear wave equation in 3-dimensional Minkowski space-time
(\ref{eq:field3}), the solution of which can be written in the integral form
\begin{eqnarray} \psi(T,x,y)&=&\frac{1}{2\pi}\frac{\partial}{\partial T}\left\{
\int_{S(T)}\frac{\psi_0(x^\prime,y^\prime)}{[T^2-(x-x^\prime)^2-(y-y^\prime)^2]^{1/2}}\,dx^\prime dy^\prime\right\}\nonumber\\
&&+\frac{1}{2\pi}\int_{S(T)}\frac{\psi_1(x^\prime,y^\prime)}{[T^2-(x-x^\prime)^2-(y-y^\prime)^2]^{1/2}}\,dx^\prime
dy^\prime,\label{intsol}\end{eqnarray} where
\[ S(T)=\{(T,x,y): T^2\geq (x-x^\prime)^2+(y-y^\prime)^2\},\]
and \[ \psi_0:= \psi|_{T=0},\quad \psi_1={\frac{\partial \psi}{\partial T}}|_{T=0}
\] are Cauchy initial data set on an arbitrary initial time slice
(which we label as $T=0$). We assume that these initial data are finite in an appropriate
sense. Imposing smoothness and compact support are sufficient, although more general data
would also satisfy our requirements \cite{abs}. This forms part of our assumption that
all initial data for the problem are regular. Then the solution (\ref{intsol}) obeys an a
priori bound which holds for all finite $T>0$ \cite{berg1,abs}. Hence for any $T_1>0$
\[ |\psi(R,T_1)|< +\infty,\quad R\geq 0.\]
So if $T_*<+\infty$, the limit equation (\ref{mainlim}) cannot be satisfied.

A similar conclusion holds in the case that $T_0=+\infty$. We can expand (\ref{intsol})
in inverse powers of $T$ to obtain a uniformly convergent series representation
\cite{abs}
\[ \psi(R,T) = \sum_{k=1}^\infty \frac{\psi_k(R)}{T^k},\]
which yields $\lim_{T\to\infty}\psi(R,T)=0$ uniformly in $R$ for all $R\geq 0$. Hence
(\ref{mainlim}) cannot be satisfied in this case, and so metric matching is ruled out.
\section{Null expansions}
We can also apply to the cylindrical case a result of Fayos, Senovilla and Torres
\cite{genmat}, that if we have two $C^3$, orientable, space-times, $\mathcal{U}^-$ and
$\mathcal{U}^+$ carrying $C^2$ metrics $g^-$ and $g^+$ respectively, then every quantity
in the resultant matched space-time $\mathcal{U}^4$ constructed from the metric, its
first derivatives and some $C^1$ tensor fields must be continuous across the boundary. In
the spherically symmetric case the null geodesic congruences are invariantly defined so
the signs of the expansion scalars of these congruences must be continuous across the
boundary.\\The ingoing radial null geodesics of the interior space-time $\mathcal{V}_-$
are given by
\[l^\mu_-=\frac{1}{a(t)}\delta^\mu_t+\frac{1}{a(t)^2}\delta^\mu_r,\]
with expansion scalar
\begin{eqnarray}
\theta_-=\nabla_{\mu}{l}_-^{\mu}=\frac{2\dot{a}}{a^2}+\frac{1}{\rho{a^2}}.
\end{eqnarray}
The ingoing radial null geodesics of the exterior space-time $\mathcal{V}^+$ have
expansion scalar
\begin{eqnarray}
\theta_-={b(U)\over{2R}}{e}^{-2(\psi-\gamma)}\quad\mbox{for some}\quad{b(U)}>0.
\end{eqnarray}
We can conclude that, in $\mathcal{V}^-$, due to the t-dependance in the scale factor\\
$a(t)=a_0|t|^{2/3}$ as $-t\rightarrow{0}$, $\theta_-\rightarrow{-}\infty$ whereas
$\mathcal{V}^+$ has $\theta_-$ strictly positive. The discontinuity in the sign of
$\theta$ across the boundary is in agreement with \cite{genmat}.  We could equivalently
show that the region $\mathcal{V}^-$ does display the formation of trapped surfaces
$\theta=\theta_+\theta_->0$ whereas the region $\mathcal{V}^+$ does not.
\section{Conclusions and Discussion}
We summarize the above as follows:
\begin{prop}
Let $(\mathcal{V}^+,g^+)$ be a vacuum cylindrically symmetric space-time with metric
described by (\ref{vacmetric}), and with the following assumptions:
\renewcommand{\theenumi}{\roman{enumi}}\begin{enumerate}
\item  In $\mathcal{V}^+$ the metric function $\omega$ is analytic,
\item  In $\mathcal{V}^+$ the metric function $\psi$ has regular initial data,
\item  $\mathcal{V}^+$ contains no trapped surfaces initially and is not radially closed.
\end{enumerate}
Let $(\mathcal{V}^-,g^-)$ be a Robertson Walker space-time with the energy conditions
$\rho>{0}$ and $\rho+p\geq{0}$. Let $(\mathcal{V}^+,g^+)$ and $(\mathcal{V}^-,g^-)$ be
matched across a $C^2$ hypersurface $\Sigma$ with continuous normal $\vec{n}$ pointing
out of $\mathcal{V}^-$ and into $\mathcal{V}^+$. Then at some value of the cosmological
time and for all subsequent times the matching breaks down.
\end{prop}
This result demonstrates the impossibility of the existence of an isotropic,
cylindrically symmetric star, that evolves from a regular initial state (or even a star
with a cylindrically symmetric portion). Matching may be possible up until a trapped
surface forms in ${\mathcal{V}^-}$ at a time $t=t^*$. By rearrangement of the matching
conditions, we can show that $t^*$ is given by the largest value of $t$ for which
\[({R_0}^{\prime})^2<1,\]where
 \[({R_0}^{\prime})^2\seq{4}{\dot{a}^2}\left(\frac{\Upsilon}{\Upsilon_{,\rho}}\right)^2.\] In the time up until $t^*$,
matching of the two space-times may be possible. However the initial conditions
necessarily imply evolution to a state where matching is not possible.

A spherically symmetric static vacuole in a dust RW cosmology was shown to be possible
\cite{ein1} and it was deduced that the observed cosmological expansion would not affect
local physics on astrophysical scales. However it then became clear that a more general
model was needed, other geometries for the interior region, and especially for the shape
of the boundary, should be considered. Senovilla and Vera \cite{straus} proved that
embedding a cylindrically symmetric \emph{static} region in an expanding RW cosmology is
always impossible irrespective of the matter inside the cavity. Mars
\cite{mars1},\cite{mars2} investigated the Einstein-Straus model with a general static
cavity embedded in a RW cosmology and obtained the result that the boundary of the static
region must be a 2-sphere and that for various reasonable energy momentum tensors the
interior is also spherically symmetric. We can consider the complementary matching of a
cylindrically symmetric vacuum interior with a RW exterior and impose regularity on the
axis of a vacuum interior without affecting the matching, i.e. the axis is not singular.
Matching of these two space-times may be possible for a finite amount of time up until a
trapped cylinder appears in the RW exterior. This leads to a contradiction and prevents
the matching from persisting and so again, we do not have a valid physical configuration.
Since our results also hold taking the vacuum region to be the interior and the RW the
exterior, they complement \cite{straus} by also
ruling out a dynamical cylindrically symmetric \emph{vacuum} interior. \\
In light of these results the impossibility of a cylindrical isotropic star is perhaps
unsurprising. However the purpose of this study is to obtain a clearer picture of simple
non-spherical, and more specifically, cylindrically symmetric systems in General
Relativity. In ongoing work, we are studying cylindrical collapse for other matter
models.

\section*{Acknowledgement} The authors wish to thank Filipe Mena for discussions and Raul Vera for his careful
reading of this paper and for his very helpful and enlightening comments. This research
is supported by Enterprise Ireland grant SC/2001/1999.
\appendix
\section{}
The $G^3_4$ field equation component for unpolarized vacuum space-time (\ref{vacmetric}):
\begin{eqnarray}\label{eq:field}
\ddel{\omega}{R}-\ddel{\omega}{T}&=&\frac{3\omega{e}^{4\psi}}{2{\alpha}^2}\left(\left(\del{\omega}{R}\right)^2-\left(\del{\omega}{T}\right)^2\right)
 +\frac{1}{\alpha}\del{\alpha}{R}\left(
\del{\omega}{R}-4\omega\del{\psi}{R}\right)+\nonumber\\
&&+2\omega\left(\del{\psi}{R}^2-\del{\psi}{T}^2 + \ddel{\gamma}{R}-\ddel{\gamma}{T}
-2\ddel{\psi}{R}+2\ddel{\psi}{T}\right)+\nonumber\\
&&+\frac{4\omega}{\alpha}\left(\del{\alpha}{T}\del{\psi}{T}\right)-
4\left(\del{\omega}{R}\del{\psi}{R}-\del{\omega}{T}\del{\psi}{T}\right)-\frac{1}{\alpha}\del{\alpha}{T}\del{\omega}{T}.
\end{eqnarray}
Field equations for polarized vacuum space-time $\alpha\neq{R}$ (\ref{vacmetric1}):
\begin{eqnarray}
\label{eq:fieldalpha1}\ddel{\alpha}{R}-\ddel{\alpha}{T}&=&0,\\
\label{eq:fieldalpha2}\del{\alpha}{R}\del{\psi}{R}-\del{\alpha}{T}\del{\psi}{T}+\alpha\left(\ddel{\psi}{R}-\ddel{\psi}{T}\right)&=&0,\\
\left(\del{\psi}{R}\right)^2 -\left(\del{\psi}{T}\right)^2+\ddel{\gamma}{R}-\ddel{\gamma}{T}&=&0,\\
\del{\gamma}{R}\del{\alpha}{T}+\del{\alpha}{R}\del{\gamma}{T}+\mixdel{\alpha}{T}{R}\left(\left(\del{\alpha}{T}\right)^2-\left(\del{\alpha}{R}\right)^2\right)&=&0.
\end{eqnarray}

Field equations for polarized vacuum space-time $\alpha={R}$ (\ref{vacmetric3}):
\begin{eqnarray}
\label{eq:field3}\frac{\partial^2\psi}{\partial{T}^2}-\frac{1}{R}\frac{\partial\psi}{\partial{R}}
-\frac{\partial^2\psi}{\partial{R}^2}&=&0,\\
\del{\gamma}{T}-2R\frac{\partial\psi}{\partial{R}}
\frac{\partial\psi}{\partial{T}}&=&0,\\
\frac{\partial\gamma}{\partial{R}}-R\left(\left(\frac{\partial\psi}{\partial{T}}\right)^2
+\left(\frac{\partial\psi}{\partial{R}}\right)^2\right)&=&0,\\
\ddel{\gamma}{R}-\ddel{\gamma}{T}-\left(\del{\psi}{T}\right)^2+\left(\del{\psi}{R}\right)^2&=&0.
\end{eqnarray}

\end{document}